\documentclass[]{spie}  
\usepackage[]{graphicx}
\usepackage{algorithmic}
\usepackage{algorithm}
\usepackage{subfigure}
\usepackage{graphicx}
\usepackage{epsfig}
\usepackage{cite}
\usepackage{threeparttable}
\usepackage{ntheorem}
\usepackage{color}
\usepackage{pdfpages}
\usepackage{authblk}

\theoremstyle{plain}
\theoremheaderfont{\normalfont\bfseries}\theorembodyfont{\slshape}
\theoremsymbol{\ensuremath{\diamondsuit}} \theoremseparator{:}
\newtheorem{mytheorem}{Theorem}

\theoremsymbol{\ensuremath{\heartsuit}}
\theoremnumbering{arabic}
\newtheorem{mylemma}{Lemma}

\theoremstyle{change}
\theorembodyfont{\upshape}
\theoremsymbol{\ensuremath{\ast}}
\theoremseparator{}

\theoremheaderfont{\sc}\theorembodyfont{\upshape}
\theoremstyle{nonumberplain} \theoremseparator{}
\theoremsymbol{\rule{1ex}{1ex}}

\theoremsymbol{\ensuremath{\heartsuit}}
\theoremnumbering{arabic}
\newtheorem{mydefinition}{Definition}

\theoremstyle{plain} \theoremsymbol{\ensuremath{\clubsuit}}
\theoremseparator{.}
\theoremnumbering{arabic}

\theoremstyle{plain} \theoremsymbol{\ensuremath{\clubsuit}}
\theoremseparator{.}
\theoremnumbering{Greek}

\theoremstyle{plain} \theoremsymbol{\ensuremath{\clubsuit}}
\theoremseparator{.}
\theoremnumbering{arabic}

\title{Self-aligned Double Patterning Friendly Configuration for Standard Cell Library Considering Placement Impact} 

\author[ ]{Jhih-Rong Gao}
\author[ ]{Bei Yu}
\author[*]{Ru Huang}
\author[ ]{David Z. Pan}
\affil[ ]{ECE Dept. Univ. of Texas at Austin, Austin, TX USA 78712}
\affil[*]{Institute of Microelectronics, Peking University, Beijing, China 100871}
\affil[ ]{\{jrgao, bei, dpan\}@cerc.utexas.edu}

\begin{document} 
  \maketitle 

\begin{abstract}
Self-aligned double patterning (SADP) has become a promising technique to push pattern resolution limit to sub-22nm technology node. Although SADP provides good overlay controllability, it encounters many challenges in physical design stages to obtain conflict-free layout decomposition. In this paper, we study the impact on placement by different standard cell layout decomposition strategies. We propose a SADP friendly standard cell configuration which provides pre-coloring results for standard cells. These configurations are brought into the placement stage to help ensure layout decomposability and save the extra effort for solving conflicts in later stages.

\end{abstract}


\section{Introduction}
Self-aligned double patterning (SADP) has become a promising technique to push pattern resolution limit to sub-22nm technology node. Compared with Litho-Etch-Litho-Etch (LELE) double patterning, SADP provides better overlay controllability with its sidewall spacer based manufacturing process. However, SADP is less mature and still exists many challenges in physical design stages. 

The most important issue for double patterning technique is how to successfully decompose the layout into two masks that are allowed to be manufactured under current 193nm optical lithography. In general, when the distance between two patterns is less than the lithography threshold, the patterns have to be separated into different masks. This process is so-called layout decomposition, or coloring. Besides of taking care of layout decomposability, overlay error is another important issue affecting manufacturability. In SADP process, the trim mask can be used to retain the desired layout region, which provides good flexibility for layout decomposition. The downside, however, is the increase of possible overlay error. Therefore, we not only need to obtain a decomposable layout, but also should seek for the layout with minimal overlay error.

Several studies \cite{yongchan_ban_flexible_2011, zhang_self-aligned_2011, xiao_polynomial_2012} have been proposed to solve layout decomposition problem for SADP. To ease the difficulty to perform layout decomposition when the layout is fixed, SADP decomposability has been considered in the routing stage \cite{mirsaeedi_self-aligned_2011, gao_flexible_2012, kodama_aspdac13}. Recently, SADP compliant designs are discussed \cite{luk-pat_design_2012, ma_self-aligned_2012} and those studies have shown how proper physical design can help the layout decomposition. Industries have been discussed the demand to integrate layout decomposition information into the placement stage. There are previous studies \cite{hsu_tcad_2011, Liebmann_spie11} focused on generating double patterning friendly standard cell library that helps to achieve layout decomposability after placement. They suggest restrictive design rules for standard cell design to ensure decomposability, such as pre-assign color for power/ground net, force single color on each cell boundary, and remove the color dependency between power/ground net and signal nets. Based on a given standard cell library, Wassala et al. \cite{Wassala_spie12} proposed an approach to find all possible combinations of cell decompositions. The techniques in these previous works involve pattern splitting and thus cannot fully apply for SADP process. Moreover, none of these works have evaluated the real impact of applying  double patterning friendly standard cells in placement.

\begin{figure}[t]
 \centering
\mbox{
\subfigure[\small{}]{\psfig{figure=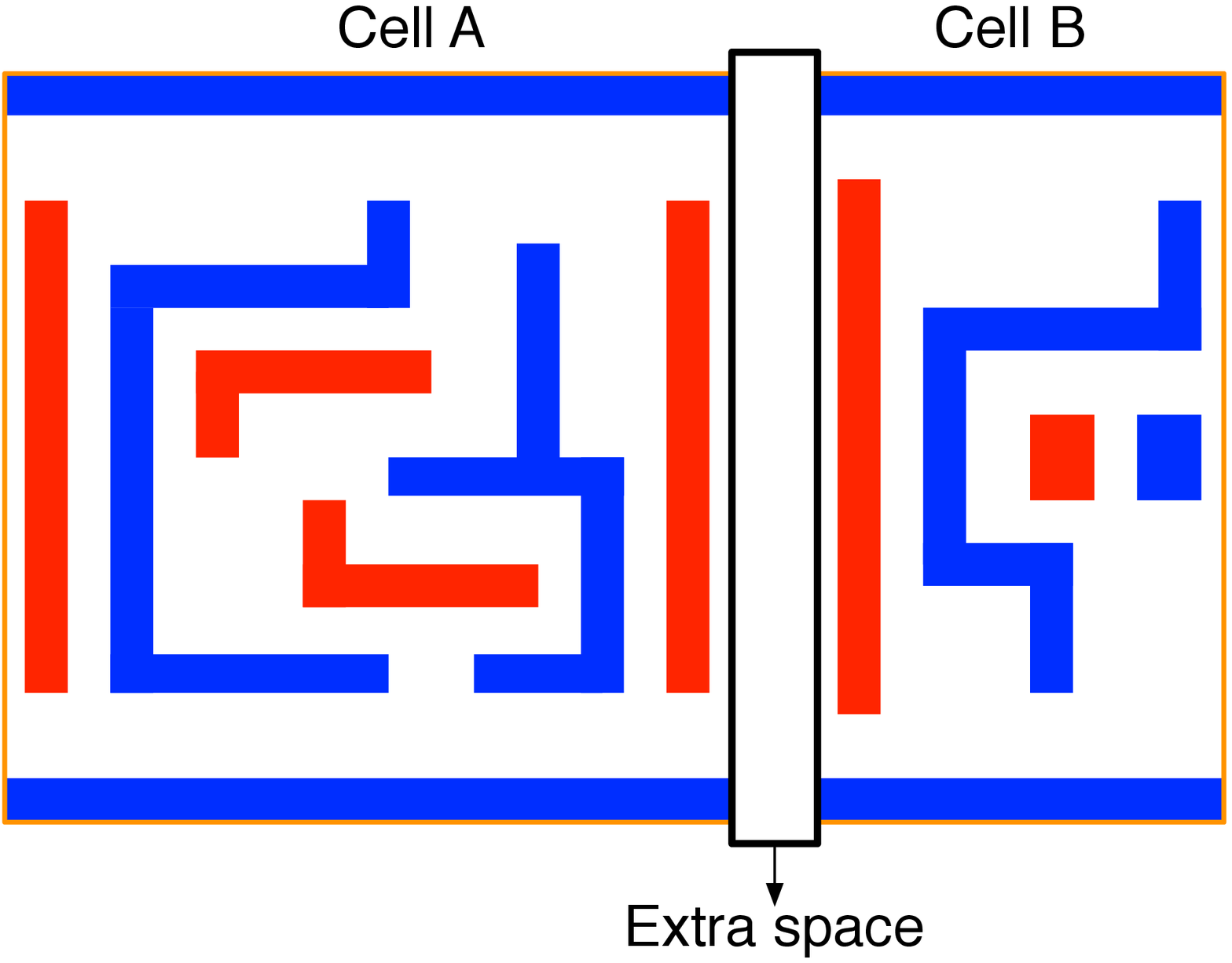,  height = 2 in}}
\hspace{0.20in}
\subfigure[\small{}]{\psfig{figure=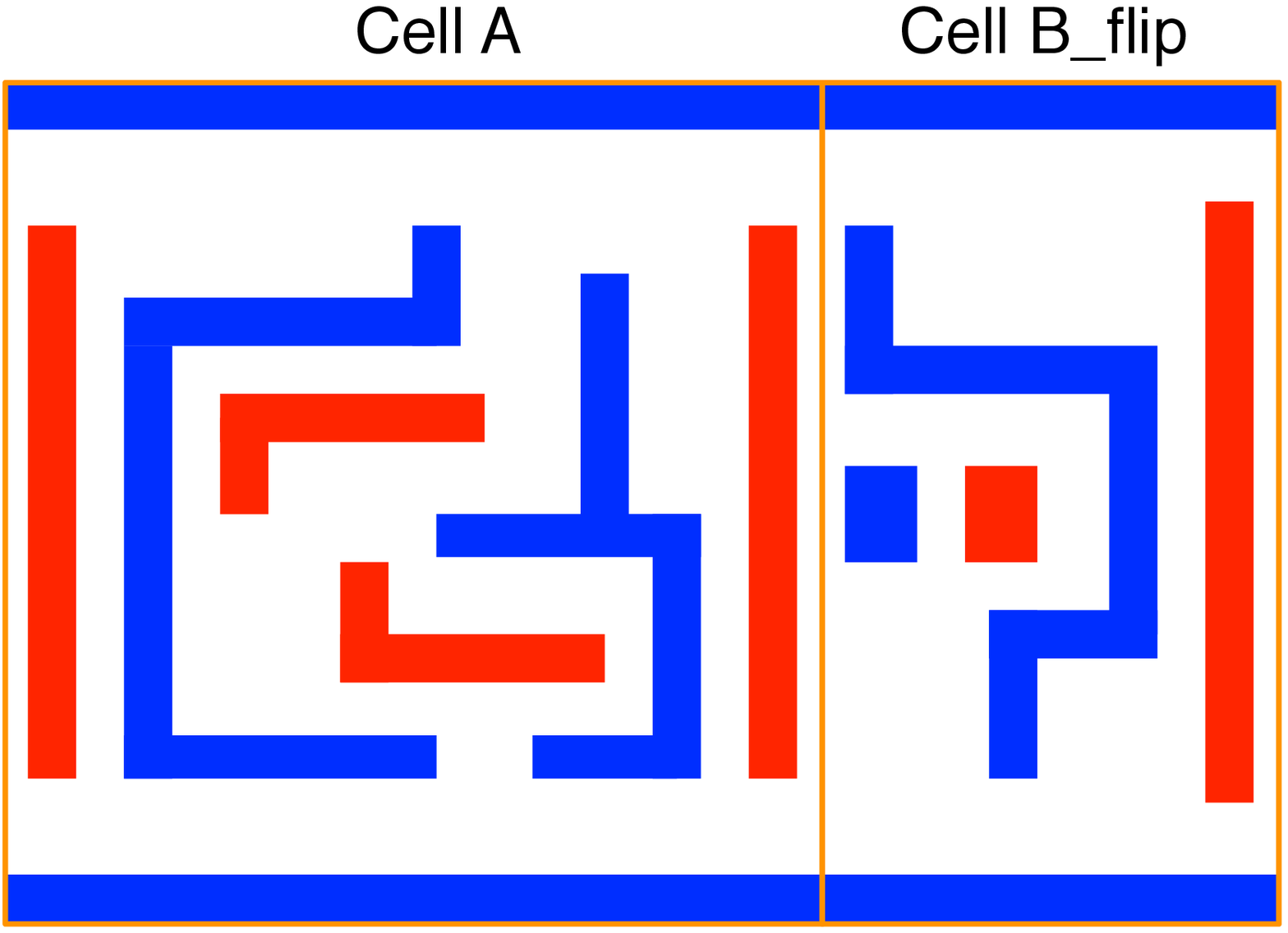,  height = 2 in}}
}
  \caption{Standard cell placement examples. (a) Improper placement requires extra space to solve the coloring conflict in the boundary. (b) SADP compliant placement is obtained by flipping Cell $B$.}
  \label{fig:place_ex}
\end{figure}

In current standard cell design methodology, designs are constructed with standard cells placed side by side in placement stage. Usually, layout decomposition is performed after place and route. However, there may be potential conflicts that need to be solved by shifting layout polygons, which may increase the layout area. By brining double patterning aware information to the placement stage, it is possible to reduce the layout area and ease the burden for fixing conflicts in later stages. Figure \ref{fig:place_ex} shows two examples of placing two standard cells, where (a) shows an improper cell placement that generates a coloring conflict by the two red patterns on the cell boundary and needs extra space between cells to resolve the conflict; while (b) shows a more compact and conflict-free placement by simply flipping Cell $B$.


In this paper, we propose a SADP friendly configuration for standard cell library considering placement impact. Our objective is to generate available configurations for standard cells, which can be used directly in the placement stage. The main contributions of the paper include: (1) we pre-define coloring results in standard cell level, and thus they can be directly applied in the placement stage. This pre-define coloring requires only one time effort, which avoids repeated layout decomposition time; 
(2) we embed the cell decomposition information into the detailed placement as SADP legalization, which helps to minimize the area impact and layout perturbation for enabling SADP; and (3) we ensure layout decomposability in an early design stage, reducing the extra effort and difficulties for fixing decomposition conflicts in later stages.


In the rest of this paper, we will discuss the impact of standard cell placement in Section \ref{sec:impact}. Our SASP-aware legalization will be explained in Section \ref{sec:method_lut}. We will discuss our experimental results in Section \ref{sec:result}, followed by the conclusion in Section \ref{sec:conclude}.

\section{Impact of standard cell placement} \label{sec:impact}
\subsection{The turnaround to placement stage}
\begin{figure}[h]
 \centering
\includegraphics[width=4in]{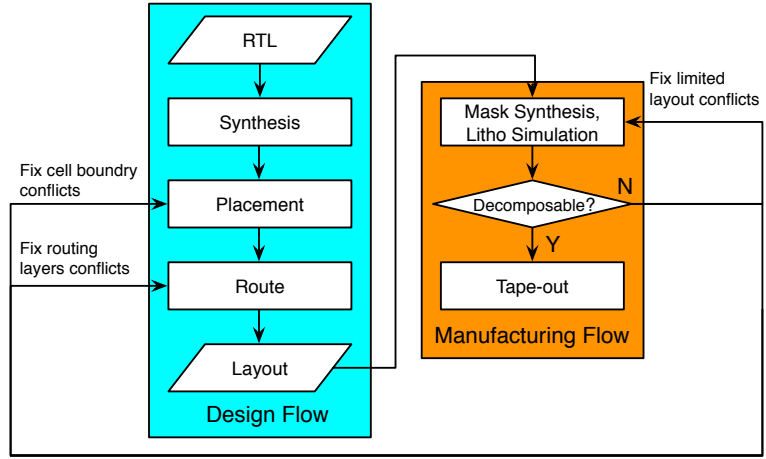}
  \caption{Design and manufacturing flow to enable deomposability.}
  \label{fig:design_flow}
\end{figure}

Double pattering lithography enables further feature shrinking to sub-22nm technology. However, there exists a gap between the product of  the design flow and a manufacturable layout. SADP process requires two adjacent patterns not fabricated on the same mask if the distance between them is less than the lithography resolution $S_{dp}$. Conflicts occur if the layout decomposition step fails to obey the spacing rule $S_{dp}$ for all patterns on the same mask. Since the general physical design flow does not take SADP awareness into consideration, it is obvious that the layout after physical design flow may not be decomposable to enable SADP. Therefore, fixing loops as shown in Figure \ref{fig:design_flow} have to be iterated until the layout is ready for manufacturing. 

Minor layout conflicts can possibly be fixed by layout decomposition techniques in mask synthesis stage. For example, SADP allows merging two conflicting patterns followed by line-cutting \cite{oyama_important_2010, lam_e-beam_2011} to generate the target layout. However, these kind of merging may be limited by restricted manufacturing rules. If there are unsolvable conflicts left, the fixing process has to go back to the routing stage to perform rip-up and re-route. Moreover, abutted placed cells may cause conflicts by patterns near the boundary. Standard cells are designed with various performance considerations, and have pre-defined pin locations. Although layout modification technique \cite{kun_yuan_wisdom:_2010, hsu_tcad_2011} can be applied to separate those closed patterns, the moving space inside a cell is limited and the performance impact is questionable. Therefore, a larger loop back to the placement stage would be necessary to fix conflicts caused between cell boundaries. The turnaround time to fix coloring conflicts can be huge if the decomposability issue is not addressed in the design flow.

There are two directions to integrate SADP awareness into the placement stage: pre-defined coloring and on-the-fly coloring. Since the layouts of standard cells are known, we can perform layout decomposition up front and embed the coloring information in the cell library that can be used by the placer (pre-defined coloring). The alternative is to perform coloring when cells are placed (on-the-fly coloring). This approach may provide the most comprehensive coloring choices, but the repeated coloring would be time-consuming. In the following paper, we will focus on the approach based on pre-defined coloring.

{\color{red}
}

\subsection{SADP challenges for standard cells}
\subsubsection{Restrictive decomposition}
\begin{figure}[b]
 \centering
\mbox{
\subfigure[\small{}]{\psfig{figure=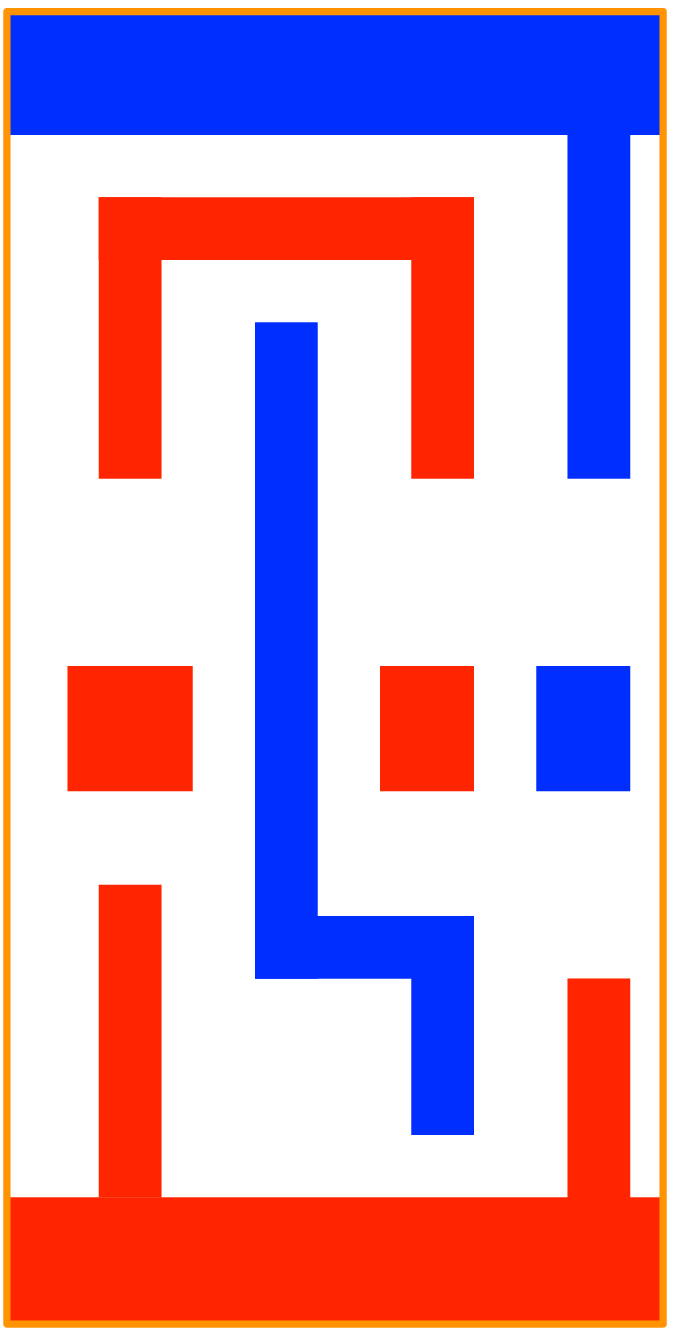,  height = 2.4 in}}
\hspace{0.20in}
\subfigure[\small{}]{\psfig{figure=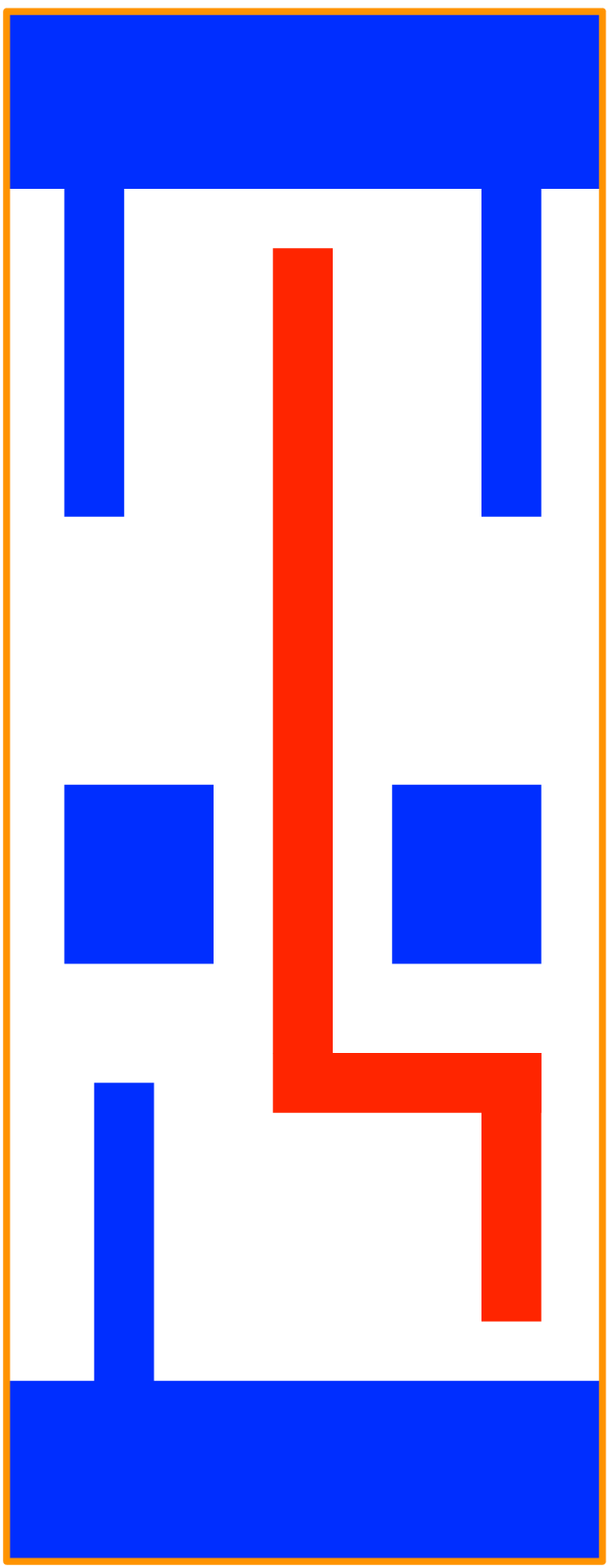,  height = 2.4 in}}
}
  \caption{Layout decomposition conflicts on power/ground net. (a) AOI21. (b) NAND2.}
  \label{fig:pg_prob}
\end{figure}

It has been mentioned that conflicts may occur in the boundary of two abutted cells. Standard cells contains power/ground net as well as signal nets, where power/ground net is connected throughout the entire row and thus is viewed as a single pattern. While conflicts between signal nets can be eliminated by simply moving the two cells apart, conflicts between power/ground net is hard to solve. Figure \ref{fig:pg_prob} shows the coloring results of two cells, where the colors of the power and ground nets are different in (a) and are the same in (b). Since the coloring results in Figure \ref{fig:pg_prob} are the only options for the two cells because of their internal pattern structures, a conflict exists natively when they are placed on the same row. For other DPL processes such as LELE-type DPL, this conflict may be solved by splitting the conflicting pattern into two. However, this technique does not work for SADP because of the process limitation.

\subsubsection{Overlay control}
\begin{figure}[h]
 \centering
\mbox{
\subfigure[\small{}]{\psfig{figure=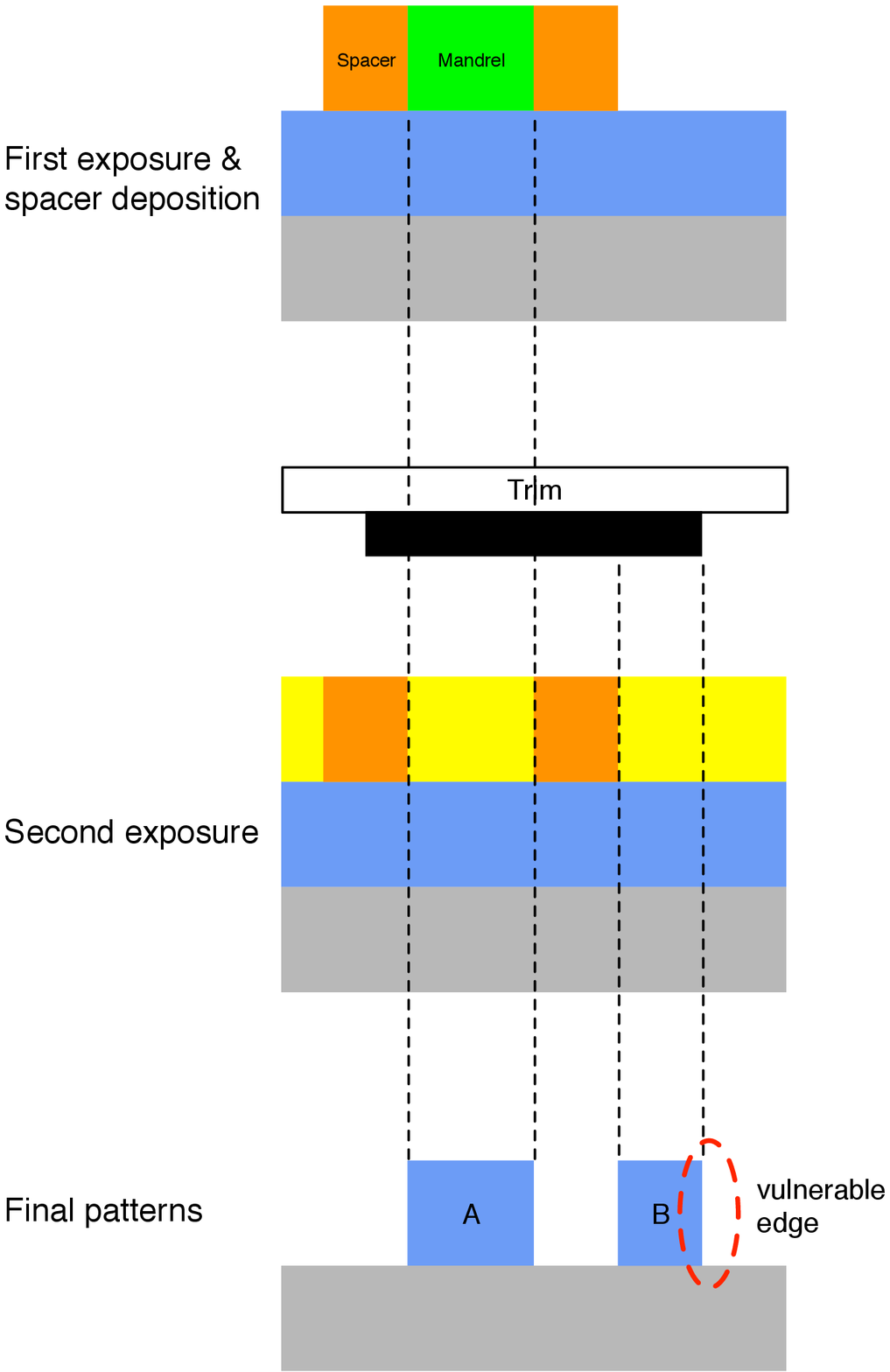,  height = 3 in}}
\hspace{0.20in}
\subfigure[\small{}]{\psfig{figure=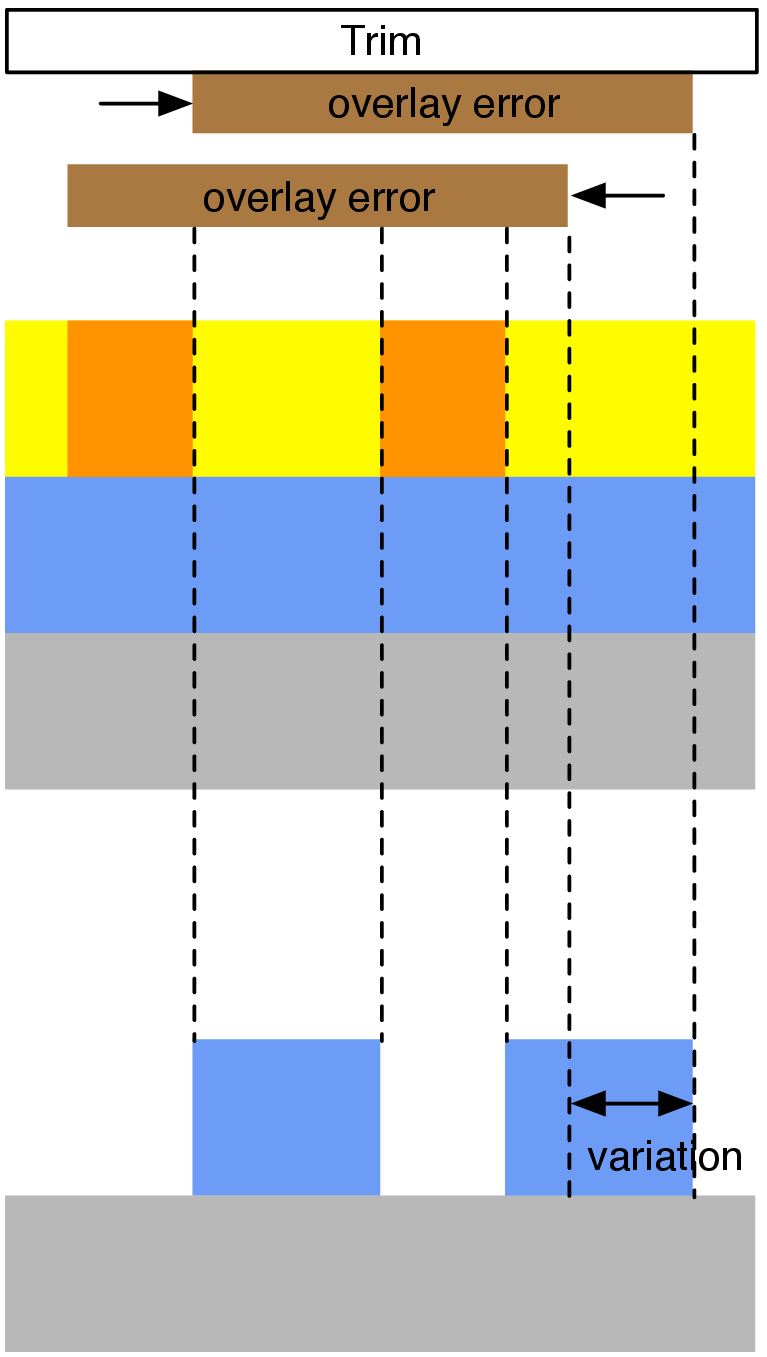,  height = 3 in}}
}
  \caption{CD variation caused by SADP overlay error.}
  \label{fig:sadp_ov}
\end{figure}

SADP process involves the first exposure with the mandrel mask, spacer deposition to the mandrels, and the second exposure with the trim mask. Overlay error may occur if two masks are not aligned perfectly. Spacers can work as  an isolating material, and thus patterns that aligned to spacers would not suffer from overlay problem. Figure \ref{fig:sadp_ov} (a) and (b) show examples with and without overlay error, respectively. Pattern $A$ is formed by the first exposure, while pattern $B$ is formed by aligning to the boundaries of the spacer and the trim pattern. If the trim mask shifts, the right edge of pattern $B$ would be vulnerable to the overlay error as shown in Figure \ref{fig:sadp_ov} (b), causing CD variation. A good layout decomposition should avoid patterns that are not aligned to spacers.

\begin{figure}[h]
 \centering
\mbox{
\subfigure[\small{}]{\psfig{figure=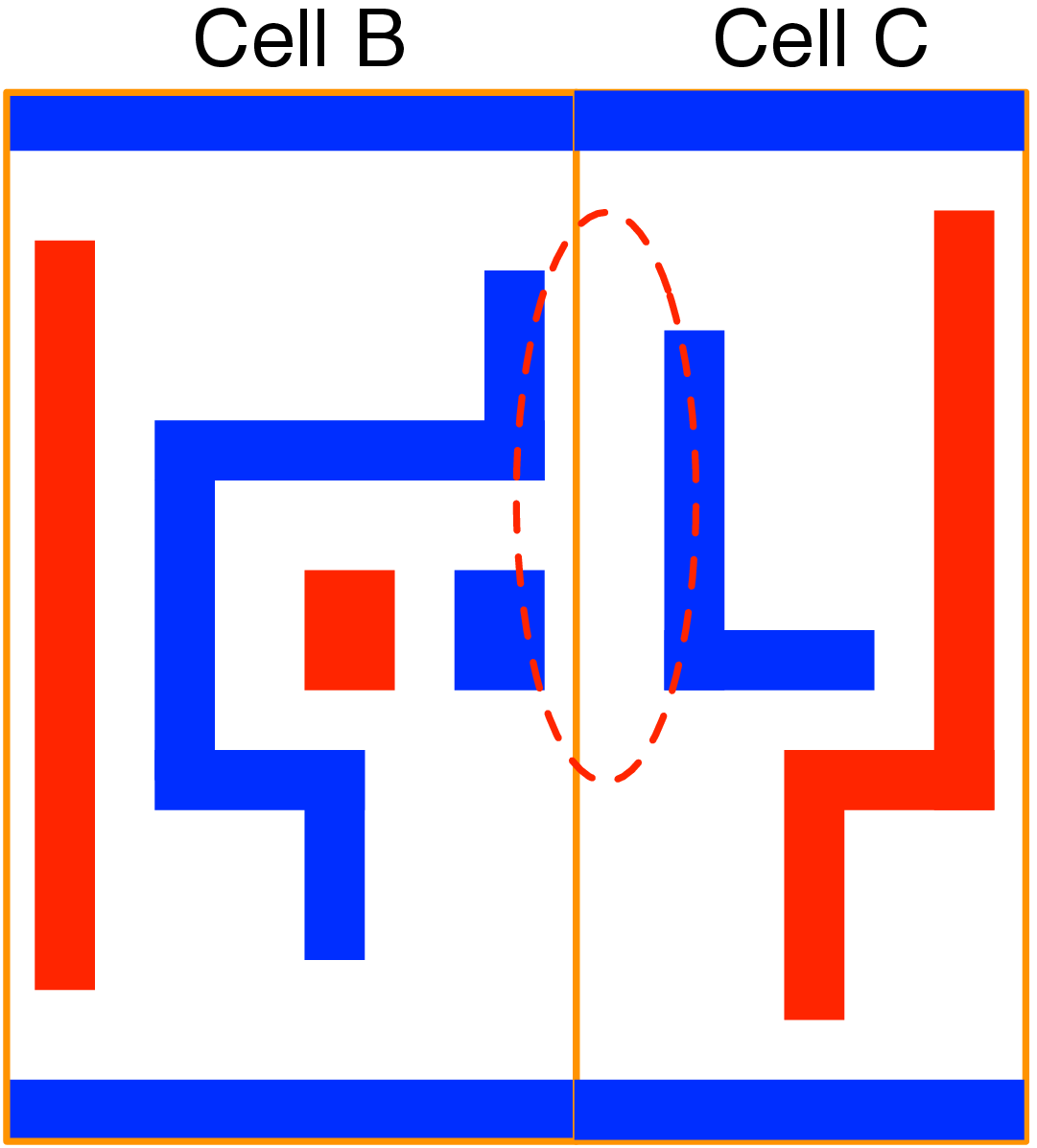,  height = 2 in}}
\hspace{0.20in}
\subfigure[\small{}]{\psfig{figure=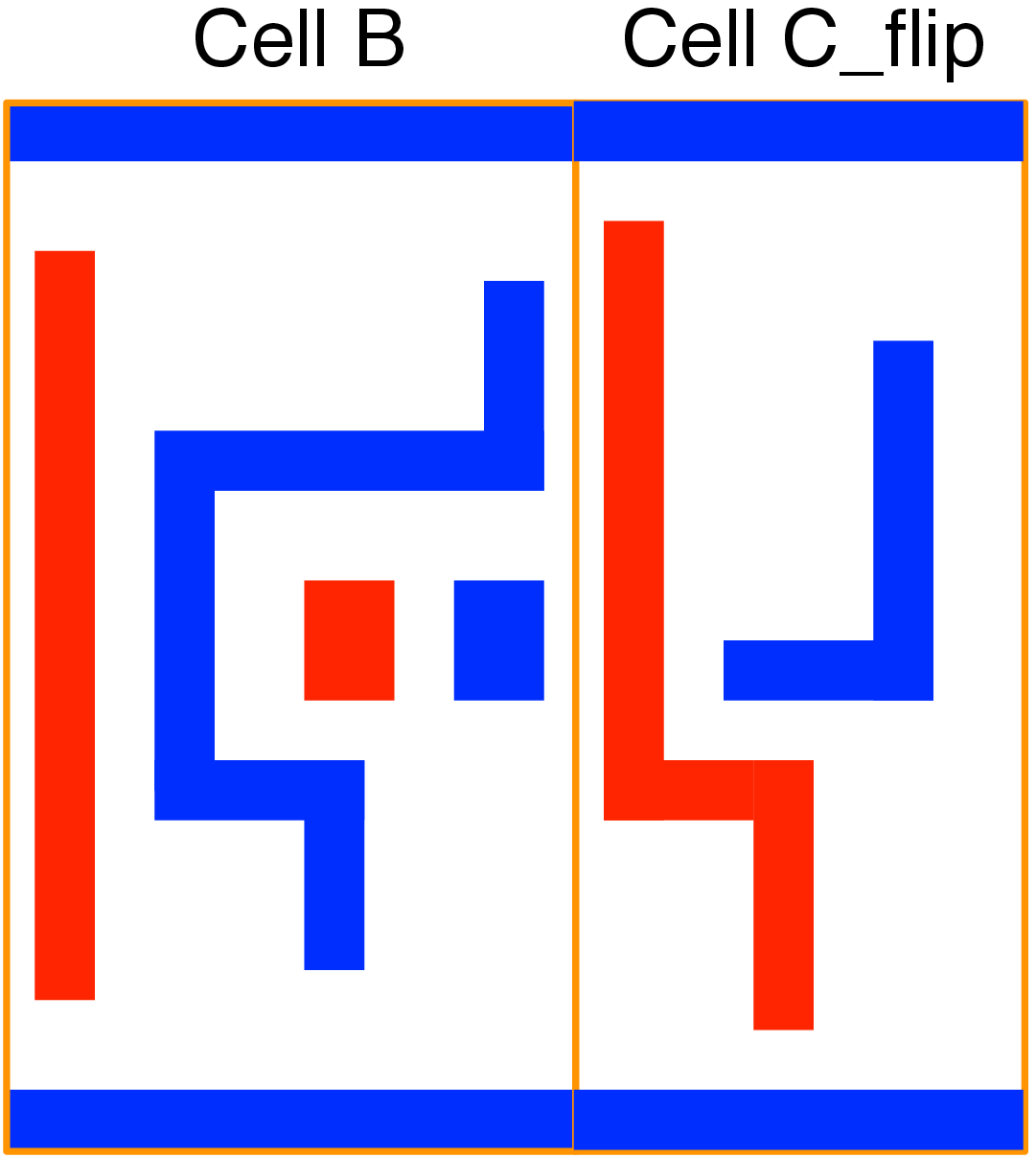,  height = 2 in}}
\hspace{0.20in}
\subfigure[\small{}]{\psfig{figure=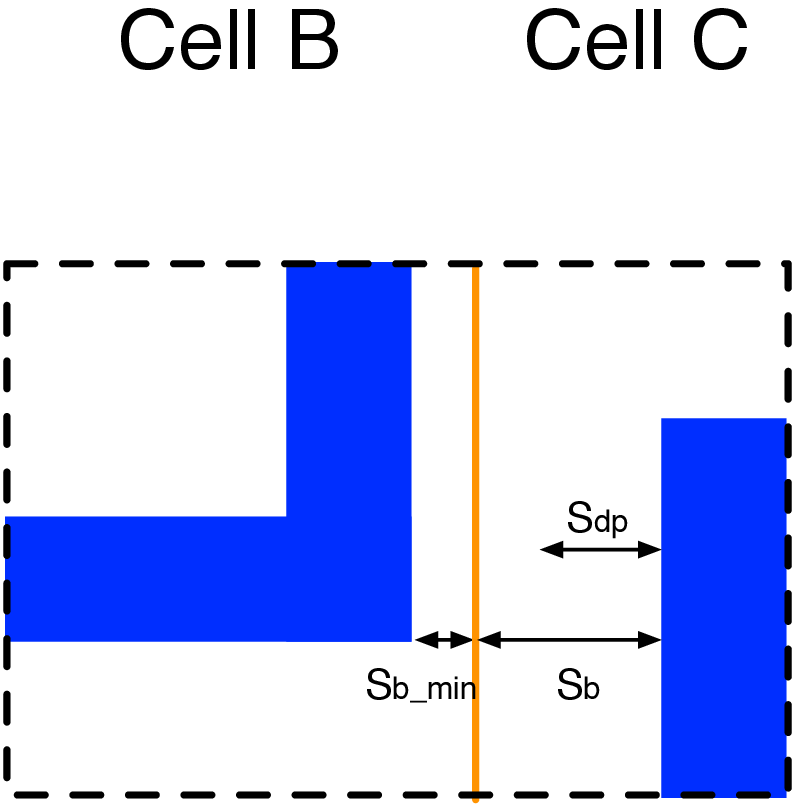,  height = 2 in}}
}
  \caption{Placement example where (a) suffers more overlay error than (b). (c) Closer view of the center area of (a).}
  \label{fig:overlay_ex}
\end{figure}

The overlay problem should be taken care when placing together two cells with pre-define colors. Figure \ref{fig:overlay_ex} shows two cases when two cells are placed, where red patterns are formed by the mandrel mask and blue pattern are formed by spacer and the trim mask. Assume there is no conflict in the boundary of cell $B$ and $C$ in (a). However, the blue patterns near the boundary would suffer from the overlay error because there is no adjacent mandrels to provide spacer alignment. Although we could reserve empty space between cell $B$ and $C$ to allow dummy mandrel insertion, this would increase the design area and wire length, and sacrifice the benefit of using DPL technology. Figure \ref{fig:overlay_ex} (b) shows the result with less overlay error by flipping cell $C$, where the mandrel in cell $C$ provides spacer alignment in the abutting boundary for cell $B$. 

\section{SADP-aware legalization} \label{sec:method_lut}
The main challenge of SADP friendly placement is the extra effort to determine the decomposability during the placement stage. In addition, the design performance would be compromised because the decomposability becomes one of the optimization objectives for the placer. The problem can be even complicated when SADP overlay minimization is also considered. Therefore, an efficient approach is needed to make SADP friendly placement possible; and in the mean while, the impact to the design performance should be minimized. We present a SADP-aware legalization applied after the regular detailed placement. We will first give the problem formulation, and discuss what SADP-aware configuration should be provided in the cell library. Then we will explain how to integrate the configuration into the placement stage efficiently.

{\color{red}
}

\subsection{Problem formulation}
Given a placed layout, our task is to identify SADP conflicts between cells, and solve them by cell flipping or cell spreading. The objective is to solve all conflicts while minimizing the core area increase and wire length perturbation. Since we only care about the conflicts between cells, we assume a cell itself is decomposable, meaning there is no conflict internally. In addition, we assume ``double-back'' rows (adjacent rows share power/ground rail) are not used in the design for more flexible decomposition results.

\subsection{Standard cell category}
\begin{figure}[b]
 \centering
\mbox{
\subfigure[\small{}]{\psfig{figure=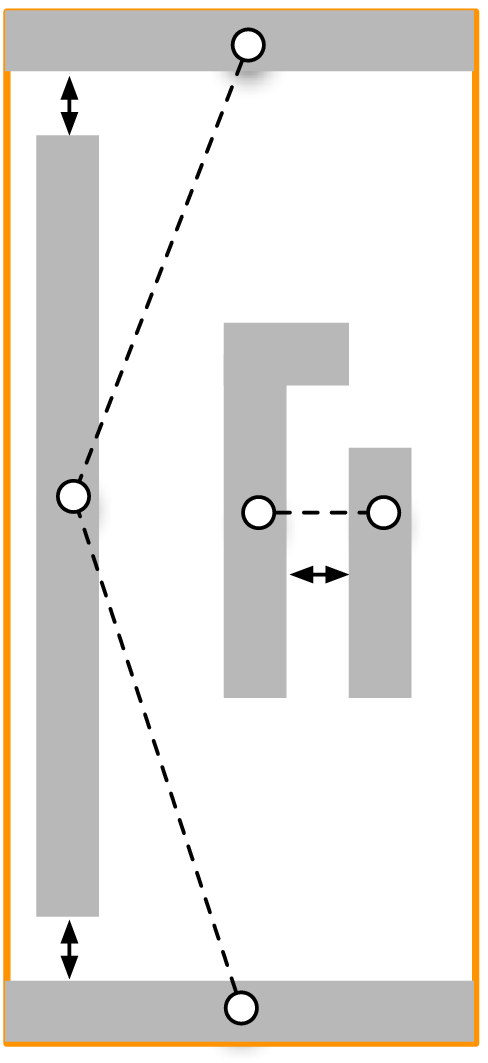,  height = 1.5 in}}
\hspace{0.20in}
\subfigure[\small{}]{\psfig{figure=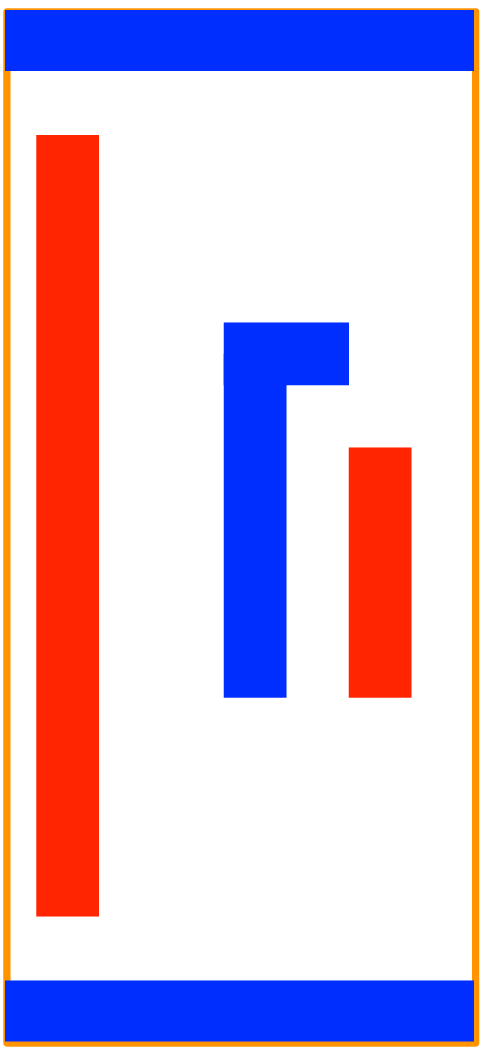,  height = 1.5 in}}
\hspace{0.20in}
\subfigure[\small{}]{\psfig{figure=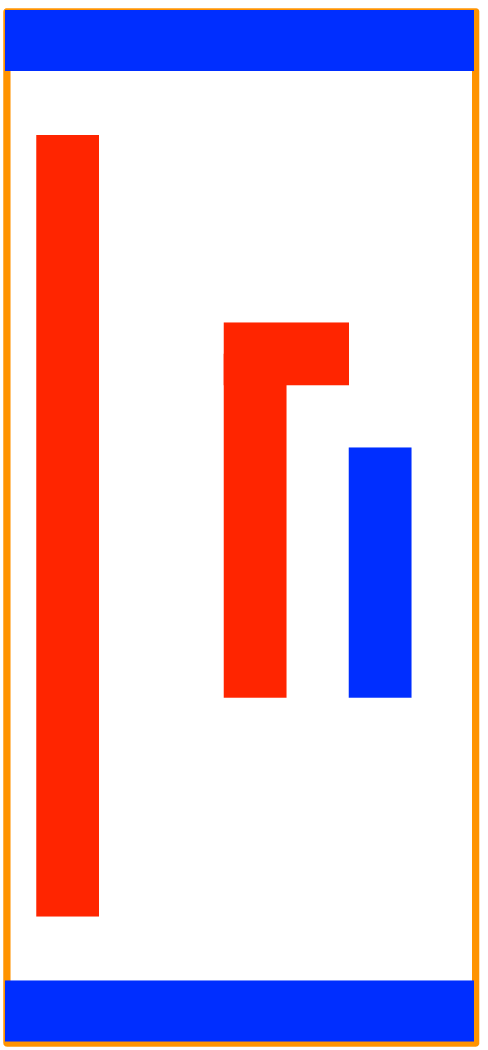,  height = 1.5 in}}
\hspace{0.20in}
\subfigure[\small{}]{\psfig{figure=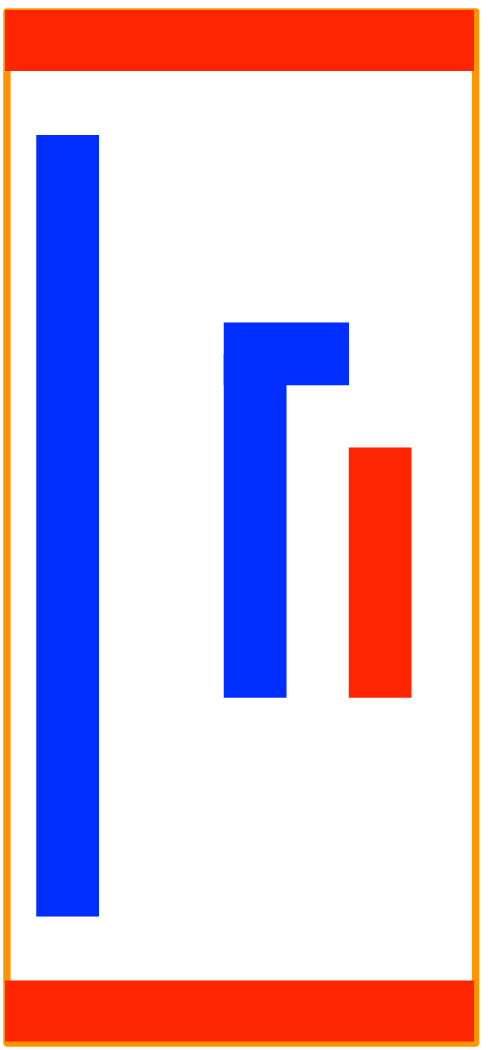,  height = 1.5 in}}
\hspace{0.20in}
\subfigure[\small{}]{\psfig{figure=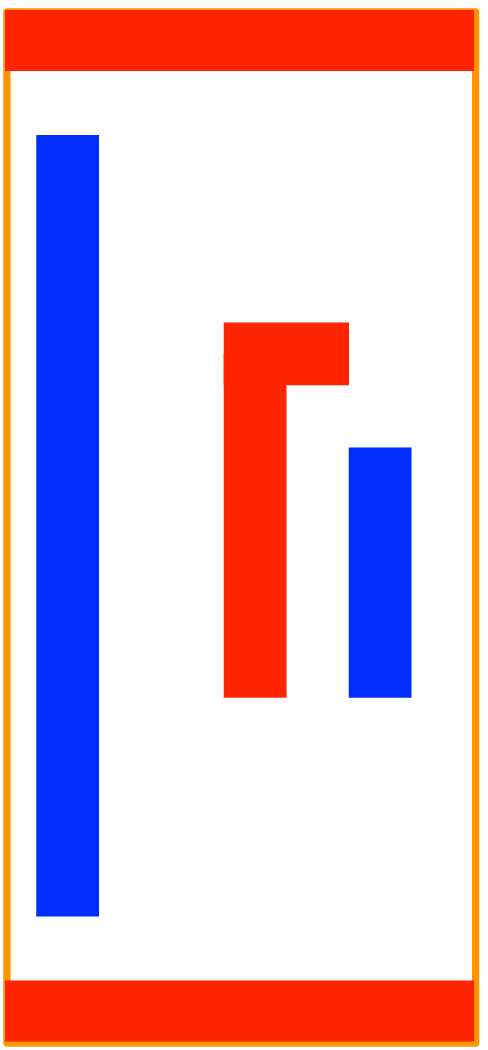,  height = 1.5 in}}
}
  \caption{(a) Sample layout and its corresponding conflict graph in dashed lines. (b)-(e) Candidate coloring results.}
  \label{fig:color_ex}
\end{figure}

Before checking and solving conflicts in the placement, we need to build pre-coloring results and embed this information in the cell library. For each cell, we build a conflict graph to represent the topological relationship among pattens. Figure \ref{fig:color_ex} (a) shows a sample layout and its corresponding conflict graph. We traverse the conflict graph in DFS manner and apply two-coloring to assign a color for each pattern. Note that the coloring assignment may not be unique. To achieve the most placement flexibility, we enumerate all possible coloring candidates for each cell as shown in Figure \ref{fig:color_ex} (b)-(e). 

Although we can always construct a conflict graph for two abutted cells and check if there is coloring conflict on the graph, this does not give much insight on how cell configurations impact decomposability. We further study the cell coloring results and categorize cells as PG-type and Abut-type which are defined as follows.

\begin{mydefinition}
PG-type: The coloring relation between power and ground nets. PG-type can be (1) Same-PG if power and ground nets must be assigned the same color; (2) Diff-PG if power and ground nets must be assigned different colors; and (3) Free-PG if there is no coloring dependency between the power and ground nets.
\end{mydefinition}

\begin{mydefinition}
Abut-type: The coloring and pattern geometry of a cell boundary (left and right). Abut-type can be (1) Safe-Abut if the spacing between the left/right-most pattern and the cell boundary is larger than the conflicting threshold $S_{th}$; (2) Free-Abut if there is no coloring dependency among the left- and right-most patterns, the power net, and the ground net; (3) Unknown-Abut if none of (1) or (2) is satisfied.
\end{mydefinition}

Examples of Diff-PG and Same-PG can be seen in Figure \ref{fig:pg_prob} (a) and (b), respectively. Free-PG exists when there is no path connecting the power and ground net on the conflict graph, so they can be assigned either the same color or different colors. Cell $A$ in Figure \ref{fig:place_ex} shows an example with Free-PG type, where a wide space between power/ground net and signal nets break their coloring dependency. It is obvious that a Same-PG cell cannot abut a Diff-PG cell; while a Free-PG cell is flexible to abut cells with any PG-type.

The left boundary of Cell $C$ in Figure \ref{fig:overlay_ex} (a) is an example of a Safe-Abut boundary. Assume there is a minimum pattern-to-boundary spacing $S_{b\_min}$ between patterns inside a cell and the cell boundary as shown in Figure \ref{fig:overlay_ex} (c). We can find a threshold distance $S_{th}$ such that as long as the pattern-to-boundary spacing $S_b$ of a cell is larger than $S_{th}$, no conflict will be induced. This threshold is determined by the resolution limit, that is, $S_b > S_{th} = S_{dp} - S_{b\_min}$. The concept of Free-Abut is similar to Free-PG, which means the coloring of patterns on the boundary are independent and have nothing to do with the power/ground net. Patterns on a Free-Abut boundary can be colored freely depending on the color of its adjacent cell. For example, the right boundary of the cell in Figure \ref{fig:color_ex} is a Free-Abut boundary, while its left boundary is Unknown-Abut because the color of the left-most patten contradicts the color of the power/ground net.

\begin{mylemma}
Two cells are PG-compatible if and only if there is no conflict between their power/ground nets, including the combinations \{Same-PG, Same-PG\}, \{Diff-PG, Diff-PG\}, and \{Free-PG, Any\}.
\end{mylemma} 

\begin{mylemma}
Two cells are Abut-compatible if and only if there is no conflict between patterns by their abutting boundary, including the combinations \{Safe-Abut, Any\}, \{Free-Abut, Any\}, and conflict-free \{Unknown-Abut, Unknown-Abut\}.
\end{mylemma} 

\begin{mytheorem} \label{the:compatible}
Two cells are compatible if and only if they are PG-compatible and Abut-compatible.
\end{mytheorem} 

We define SADP compatibility of a pair of cells as the decomposability when they are placed abutted, which can be determined by Theorem \ref{the:compatible}. Determining PG-compatibility is trivial, however, determining Abut-compatibility may requires extra effort if two Unknown-Abut boundaries are considered. In that case, we need to construct a conflict graph for patterns of two abutted cells and check if odd cycles (conflicts) are formed in the graph.

\subsection{Decomposability look-up table} \label{sec:dplut}
Although determining PG-compatibility is trivial, determining Abut-compatibility requires extra effort when Unknown-Abut boundaries involve. It is inefficient to perform coloring for Unknown-Abut boundaries whenever they are checked. Since cell library has fixed layouts and usually contains only hundreds of cells, we can create a decomposability look-up table (DPLUT) to provide quick query during legalization.

\begin{figure}[h]
 \centering
\includegraphics[width=5.5in]{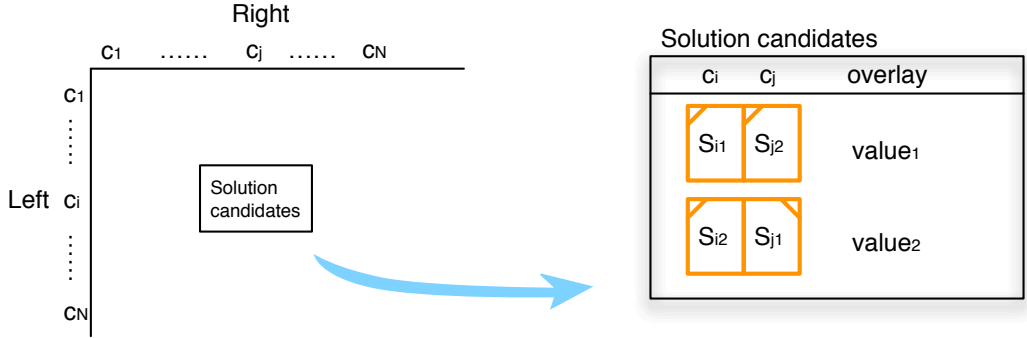}
  \caption{Decomposability look-up table.}
  \label{fig:lut}
\end{figure}

Given a standard cell library with $N$ cells, we build a two-dimensional $N \times N$ DPLUT based on Theorem \ref{the:compatible}, in which each table entry keeps the decomposable solution candidates of two abutted cells. The first dimension represents the cell on the left, and the second dimension represents the cell on the right, as shown in Figure \ref{fig:lut}. The solution candidates indicate decomposable cell orientations and the corresponding overlay error on the abutting boundary. If there is no legal combination to place two cells, their solution candidates would be $NULL$. Note that the same cell orientation may have different SADP layout decomposition results based on how patterns are colored, and we only keep the one with the minimum overlay error. With DPLUT, we can quickly query if two cells can be put together, and obtain the minimal overlay orientation as their placement solution. For example, the solution candidates in Figure \ref{fig:lut} indicate two orientations to place cell $c_i$ and $c_j$, and thus we can decide wether to perform cell flipping according to the possible overlay error $value_1$ and $value_2$. Solutions $S_{i1}$ and $S_{i2}$ represent two different coloring results for cell $c_i$.

\subsection{SADP legalization}
We present a post processing in detailed placement, SADP legalization, to ``legalize'' conflicting cells after the regular placement without significant design perturbation. After obtaining the DPLUT mentioned in Section \ref{sec:dplut}, we use it to quickly determine the decomposability between two adjacent cells. Once a conflict is found, we applied one of the two techniques, cell flipping and cell spreading, to resolve it. When the design area is the main concern, cell flipping is preferred because it does not impose area overhead. Cell spreading may take advantage of existing white space on the original placement, but the  white space may not be sufficient to solve all conflicts. In certain placement, we may need to enlarge the design area in order to resolve all conflicts.

Algorirthm \ref{alg:sadp_legal} shows our greedy-based legalization. Rows of the placement are processed one by one, left to right (Line 1, 2). For each pair of conflicting cells, we first check if the conflict can be solved by flipping one cell as shown in Line \ref{alg:flip}. If a flipped solution is not available, then we try cell spreading as shown in Line \ref{alg:spread}. Decomposability check ``IsConfclit'' is done by checking if \{$c_{r_i}$, $c_{r_{i+1}}$\} in the current orientation exists in DPLUT[$c_{r_i}$][$c_{r_{i+1}}$]. If not, we may flip one of the cells according to the solution candidates in DPLUT[$c_{r_i}$][$c_{r_{i+1}}$] and the current color assignment of the two cells. However, if there is no any way to decompose the two adjacent cells, we can only solve their conflict by spreading the two cells.

\begin{figure}[h]
\center
 \begin{minipage}[t]{3.15in}
 \begin{algorithm}[H]
\caption{SADP legalization} 
\label{alg:sadp_legal} 
\begin{algorithmic}[1]
\REQUIRE $R$ rows of cells
\FOR {each $r \in R$}
  \FOR {$i= 1 \to |r| - 1$} 
    \IF {IsConflict($c_{r_i}$, $c_{r_{i+1}}$)}
      \STATE Flip($c_{r_i}$, $c_{r_{i+1}}$) \label{alg:flip}
    \ENDIF
  \ENDFOR
  \FOR {$i= 1 \to |r| - 1$} 
    \IF {IsConflict($c_{r_i}$, $c_{r_{i+1}}$)}
      \STATE Spread($c_{r_i}$, $c_{r_{i+1}}$) \label{alg:spread}
    \ENDIF
  \ENDFOR
\ENDFOR
\end{algorithmic}
\end{algorithm}
 \end{minipage}
\end{figure}


\section{Experimental Results} \label{sec:result}
Existing placement benchmarks such as ISPD 06' benchmark only provide placement information without standard cell library detail. Therefore, those benchmarks cannot be used for SADP legalization. Instead, we synthesize OpenSPARC T1 designs with Nangate 45nm standard cell library  \cite{nangate} to generate our benchmark. For simplicity, we assume the sizes of the minimum pattern width, spacing, and spacer width are the same, and modify the layout accordingly. Cells are decomposed as explained in Section \ref{sec:method_lut} and used to configure the decomposability look-up table.
Because Nangate standard cell library is not designed for SADP, several cells are not decomposable internally. For simplicity, we assume there is no internal conflicts and no power/ground incompatibility, so we can focus on solving conflicts between cell boundaries. 

\subsection{SADP legalization}
We perform placement with Cadence SOC Encounter \cite{socEncounter} and use the result as the input of our approach. The default core utilization rate is set as $0.7$. The benchmark information and our results are shown in Table \ref{tab:res_ub}. 

We implement two versions of SADP-legalization, area-unbounded (UB) and area-bounded (B). In SADP-legalization\_UB, expanding layout area is allowed for cell spreading if necessary. Table \ref{tab:res_ub} shows the results, where all conflicts in the original placement are solved with slight area and wire length perturbation. On average, SADP-legalization\_UB induces $3.25\%$ additional area and $1.39\%$ additional wire length.

In SADP-legalization\_B shown in Table \ref{tab:res_b}, layout area is fixed, and conflicts can only be solved within the given area specification. Our results show that around 40\% of the conflicts can be solved without any area penalty, and the wire length perturbation is only 0.08\% on average. SADP-legalization for all designs can be accomplished within seconds, showing the efficiency of utilizing DPLUT.

\begin{table*}[htbp]
\caption{Experimental results with area-unbounded SADP legalization.}
\begin{center}
\begin{tabular}{c|ccc|ccccc}
\hline
\multicolumn{ 1}{c|}{Design} & \multicolumn{ 3}{c|}{Benchmark Statistics} & \multicolumn{ 5}{c}{SADP-Legalization\_UB} \\ \cline{ 2- 9}
\multicolumn{ 1}{c|}{} & Conflict & Area($um^2$) & WL($um$) & Conflict & Area($um^2$) & WL($um$) & +Area\% & +WL\% \\ \hline
alu & 877 & 5284 & 29620 & 0 & 5451 & 30004 & 3.16\% & 1.30\% \\
byp & 2089 & 18011 & 133500 & 0 & 18997 & 135635 & 5.47\% & 1.60\% \\
div & 1439 & 11860 & 55390 & 0 & 12785 & 56535 & 7.80\% & 2.07\% \\ 
ecc & 587 & 5046 & 23090 & 0 & 5225 & 23376 & 3.55\% & 1.24\% \\
ffu & 612 & 6493 & 27970 & 0 & 6564 & 28216 & 1.09\% & 0.88\% \\ 
mul & 5463 & 42139 & 205500 & 0 & 42224 & 207978 & 0.20\% & 1.21\% \\ 
efc & 454 & 4471 & 12150 & 0 & 4536 & 12326 & 1.45\% & 1.45\% \\ \hline
Average &  &  &  &  &  &  & 3.25\% & 1.39\% \\ \hline
\end{tabular}
\end{center}
\label{tab:res_ub}
\end{table*}

\begin{table}[htbp]
\caption{Experimental results with area-bounded SADP legalization.}
\begin{center}
\begin{tabular}{c|c|c|c|c}
\hline
\multicolumn{ 1}{c|}{Design} & \multicolumn{ 4}{c}{SADP-Legalization\_B} \\ \cline{ 2- 5}
\multicolumn{ 1}{c|}{} & Conflict & WL & -Conflict\% & +WL\% \\ \hline
alu & 519 & 29648 & 40.82\% & 0.09\% \\ 
byp & 1433 & 133552 & 31.40\% & 0.04\% \\ 
div & 831 & 55437 & 42.25\% & 0.08\% \\ 
ecc & 364 & 23108 & 37.99\% & 0.08\% \\ 
ffu & 345 & 27991 & 43.63\% & 0.08\% \\ 
mul & 3133 & 205683 & 42.65\% & 0.09\% \\ 
efc & 267 & 12164 & 41.19\% & 0.12\% \\ \hline
Average &  &  & 39.99\% & 0.08\% \\ \hline
\end{tabular}
\end{center}
\label{tab:res_b}
\end{table}


\subsection{Analysis of SADP-friendly standard cell design and placement}
There are two aspects when talking about SADP-friendly standard cell design. One is internally SADP-friendly, meaning the cell itself is decomposable. Another is externally SADP-friendly, meaning the cell can easily abut on another cell without conflicts. It is essential to ensure standard cells are self-decomposable to achieve basic layout decomposability. It is also important to maintain external SADP friendliness, which can further improve placement results. We analyze the layout after applying SADP-legalization, and observe some important factors that affect the result quality. Below we discuss some design strategies that can improve SADP-aware legalization. This information can be considered by cell designers and CAD engineer to better achieve SADP friendliness.

\begin{figure}[t]
 \centering
\includegraphics[width=3in]{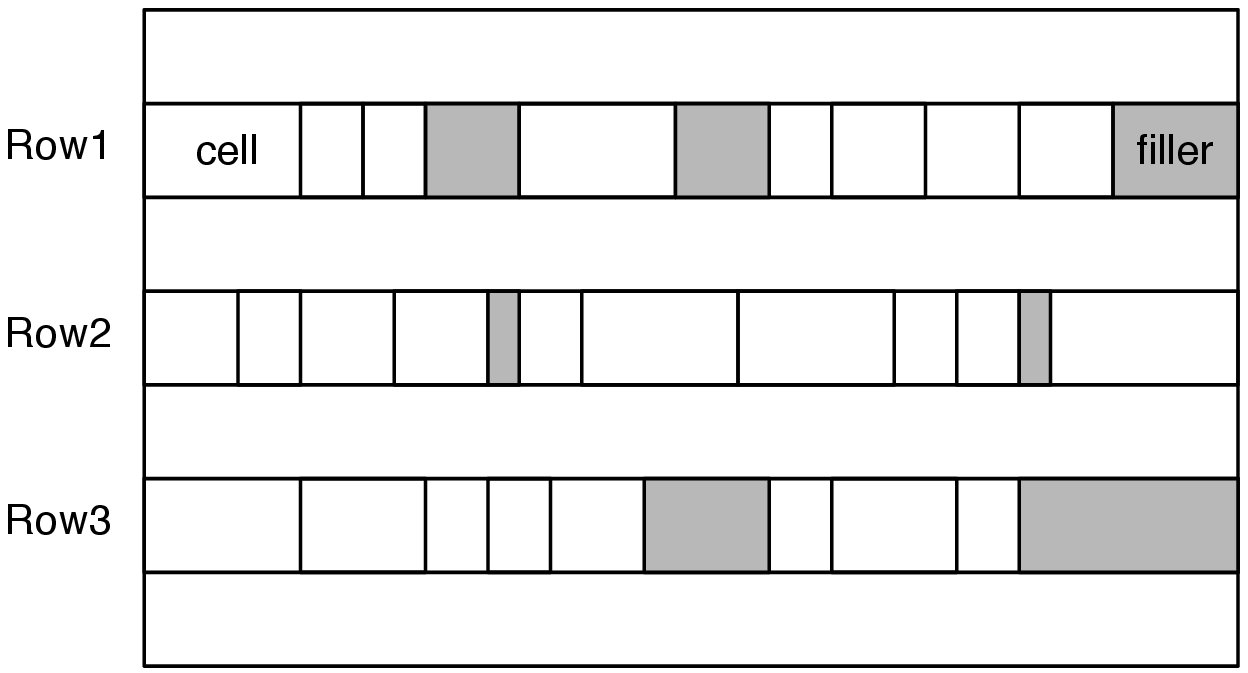}
  \caption{Row 2 may cause conflict solving bottleneck.}
  \label{fig:cong_row}
\end{figure}

\begin{itemize}
  \item The power/ground compatibility between cells is the biggest issue in SADP-aware placement. The cell library either needs to maintain consistent coloring of power/ground net for all cells, or it needs to provide both coloring options (the same or different coloring) for each cell to provide the most flexibility for decomposable placement.
  \item The patterns on the boundary of a cell is where conflicts may occur, and thus the pattern number on the boundary should be kept small and the pattern structure on the boundary should be as simple as possible. For example, if patterns on the boundary are assigned the same color, they can easily abut a cell with another color on the boundary. However, if the patterns are with mixed colors, finding the conflict-free match would be more difficult.
  \item Leaving white space would benefit SADP legalization. Most importantly, the white space should be evenly distributed among the rows or the entire layout to avoid the bottleneck of solving conflicts. The area increase or performance degradation is usually determined by the row or region that need the most white space for solving conflicts. If a particular row is much congested than others, the chance is that solving its conflicts needs to allocate more space by enlarging the core area. For example, Row 2 in Figure \ref{fig:cong_row} is more congested than the others; performing cell spreading for it requires either increasing the core area or moving cells in Row 2 to other rows with larger performance impact.
\end{itemize}

\section{Conclusions} \label{sec:conclude}
Double patterning enables nanometer lithography, however, achieving decomposable designs is still challenging. We study the coloring strategies of standard cells and analyze their impact to the placement with SADP process. We present the standard cell configuration and embed this information in the cell library. Based on the cell configuration, our SADP legalization can quickly determine the decomposability between cells and solve conflicts with cell flipping and cell spreading techniques. The results show that our approach can efficiently solve conflicts with small area overhead and layout perturbation.

\section{Acknowledgements} 
This work is supported in part by NSF, SRC, Oracle, and NSFC.

\bibliography{stdcell}   

\begin{thebibliography}{10}

\bibitem{yongchan_ban_flexible_2011}
Ban, Y., Lucas, K., and Pan, D.~Z., ``Flexible {2D} layout decomposition
  framework for spacer-type double pattering lithography,'' in {\em Proc.
  Design Automation Conf.}{\nolinebreak\hspace{0.1em}},   789--794 (2011).

\bibitem{zhang_self-aligned_2011}
Zhang, H., Du, Y., Wong, M. D.~F., and Topaloglu, R., ``Self-aligned double
  patterning decomposition for overlay minimization and hot spot detection,''
  in {\em Proc. Design Automation Conf.}{\nolinebreak\hspace{0.1em}},   71
  (2011).

\bibitem{xiao_polynomial_2012}
Xiao, Z., Du, Y., Zhang, H., and Wong, M.~D., ``A polynomial time exact
  algorithm for self-aligned double patterning layout decomposition,'' in {\em
  Proc. Int. Symp. on Physical Design}{\nolinebreak\hspace{0.1em}},   17–24,
  {ACM} (2012).

\bibitem{mirsaeedi_self-aligned_2011}
Mirsaeedi, M., Torres, J.~A., and Anis, M., ``Self-aligned double-patterning
  {(SADP)} friendly detailed routing,'' in {\em Proc. of
  SPIE}{\nolinebreak\hspace{0.1em}},   {\bf 7974},  79740O--79740O--9 (2011).

\bibitem{gao_flexible_2012}
Gao, J.-R. and Pan, D.~Z., ``Flexible self-aligned double patterning aware
  detailed routing with prescribed layout planning,'' in {\em Proc. Int. Symp.
  on Physical Design}{\nolinebreak\hspace{0.1em}},   25–32 (2012).

\bibitem{kodama_aspdac13}
Kodama, C., H., I., Nakayama, K., Kotani, T., Nojima, S., Mimotogi, S.,
  Miyamoto, S., and Takahashi, A., ``Self-aligned double and quadruple
  patterning-aware grid routing with hotspots control,'' in {\em Proc. Asia and
  South Pacific Design Automation Conf.}{\nolinebreak\hspace{0.1em}},  (2013).

\bibitem{luk-pat_design_2012}
Luk-Pat, G., Miloslavsky, A., Painter, B., Lin, L., De~Bisschop, P., and Lucas,
  K., ``Design compliance for spacer is dielectric {(SID)} patterning,'' in
  {\em Proc. of SPIE}{\nolinebreak\hspace{0.1em}},   {\bf 8326}(1),
  83260D--83260D--13 (2012).

\bibitem{ma_self-aligned_2012}
Ma, Y., Sweis, J., Yoshida, H., Wang, Y., Kye, J., and Levinson, H.~J.,
  ``Self-aligned double patterning {(SADP)} compliant design flow,'' in {\em
  Proc. of SPIE}{\nolinebreak\hspace{0.1em}},   {\bf 8327}(1),
  832706--832706--13 (2012).

\bibitem{hsu_tcad_2011}
Hsu, C.-H., Chang, Y.-W., and Nassif, S.~R., ``Simultaneous layout migration
  and decomposition for double patterning technology,'' in {\em IEEE Trans. on
  Computer-Aided Design of Integrated Circuits and
  Systems}{\nolinebreak\hspace{0.1em}},   284--294 (2011).

\bibitem{Liebmann_spie11}
Liebmann, L., Pietromonaco, D., and Graf, M., ``Decomposition-aware standard
  cell design flows to enable double-patterning technology,'' in {\em Proc. of
  SPIE}{\nolinebreak\hspace{0.1em}},   {\bf 7974},  79740K--79740K (2011).

\bibitem{Wassala_spie12}
Wassala, A.~G., Sharafb, H., and Hammouda, S., ``Placement-aware decomposition
  of a digital standard cells library for double patterning lithography,'' in
  {\em Proc. of SPIE}{\nolinebreak\hspace{0.1em}},   {\bf 8522},  852222
  (2012).

\bibitem{oyama_important_2010}
Oyama, K., Nishimura, E., Kushibiki, M., Hasebe, K., Nakajima, S., Murakami,
  H., Hara, A., Yamauchi, S., Natori, S., Yabe, K., Yamaji, T., Nakatsuji, R.,
  and Yaegashi, H., ``The important challenge to extend spacer {DP} process
  towards 22nm and beyond,'' in {\em Proc. of
  SPIE}{\nolinebreak\hspace{0.1em}},   {\bf 7639},  763907--763907--6 (Mar.
  2010).

\bibitem{lam_e-beam_2011}
Lam, D., Liu, E.~D., Smayling, M.~C., and Prescop, T., ``E-beam to complement
  optical lithography for {1D} layouts,'' in {\em Proc. of
  SPIE}{\nolinebreak\hspace{0.1em}},   {\bf 7970},  797011--797011--8 (2011).

\bibitem{kun_yuan_wisdom:_2010}
Yuan, K. and Pan, D.~Z., ``{WISDOM:} wire spreading enhanced decomposition of
  masks in double patterning lithography,'' in {\em International Conference on
  Computer-Aided Design}{\nolinebreak\hspace{0.1em}},   32--38 (Nov. 2010).

\bibitem{nangate}
``{NanGate FreePDK45 Generic Open Cell Library}.''
\newblock [http://www.si2.org/openeda.si2.org/projects/nangatelib].

\bibitem{socEncounter}
``{Cadence SOC Encounter}.''
\newblock [http://www.cadence.com/].

\end{thebibliography}
\bibliographystyle{spiebib}   

\end{document}